\documentclass[journal]{IEEEtran}
\usepackage{amssymb}
\usepackage[table]{xcolor}
\usepackage{bm}
\usepackage{amsthm}
\usepackage{amsmath}	
\usepackage{stfloats}
\usepackage{algorithm}
\usepackage{algorithmicx}
\usepackage{algpseudocode}
\usepackage{graphics}
\usepackage[demo]{graphicx}
\usepackage{multirow}
\usepackage{array}
\usepackage{subfigure}
\usepackage{setspace}
\usepackage{footnote}
\makesavenoteenv{tabular}
\makesavenoteenv{table}
\usepackage{color}
\usepackage{booktabs}
\usepackage{colortbl}

\newcommand{\reffig}[1]{Fig. \ref{#1}}

\renewcommand{\eqref}[1]{(\ref{#1})}

\newcommand{\hl}[1]{\textcolor{black}{#1}} 

\begin{document}
\title{ Artificial Intelligence Powered Mobile Networks: From Cognition to Decision }

\author{Guiyang Luo, Quan Yuan, Jinglin Li,  Shangguang Wang, and Fangchun Yang
	\thanks{This work was supported  the	National Natural Science Foundation of China  under Grant 6210070488, Grant 61902035 and Grant 61876023.  \textit{(Corresponding author: Yuan Quan.)}
	}
	\thanks{G. Luo,  Q. Yuan, J. Li, S. Wang, and  F. Yang are with the
		State Key Laboratory of Networking and Switching Technology, Beijing
		University of Posts and Telecommunications, Beijing, 100876 China (e-mail:
		\{luoguiyang,  yuanquan, jlli, sgwang,  fcyang\}@bupt.edu.cn).}
}

\maketitle

\begin{abstract}

Mobile networks (MN) are anticipated to provide unprecedented opportunities to enable a new world of connected experiences and radically shift the way people interact with everything. 
MN are becoming more and more complex, driven by ever-increasingly complicated configuration issues and blossoming new service requirements. This complexity poses significant challenges in deployment, management, operation, optimization, and maintenance, since they require a complete understanding and cognition of MN. 
Artificial intelligence (AI), which deals with the simulation of intelligent behavior in computers, has demonstrated enormous success in many application domains, suggesting its potential in cognizing the state of MN and making intelligent decisions. 
In this paper, we first propose an AI-powered mobile network architecture and discuss challenges in terms of cognition complexity, decisions with high-dimensional action space, and self-adaption to system dynamics. Then, potential solutions that are associated with AI are discussed. Finally, we propose a deep learning approach that directly maps the state of MN to perceived QoS, integrating cognition with the decision. Our proposed approach helps operators in making more intelligent decisions to guarantee QoS. Meanwhile, the effectiveness and advantages of our proposed approach are demonstrated on  a real-world dataset, involving $31261$ users over $77$ stations within $5$ days.
	
\end{abstract}

\begin{IEEEkeywords}
Artificial intelligence, Machine learning, Mobile network, Network managements
\end{IEEEkeywords}
\IEEEpeerreviewmaketitle

\section{Introduction}
The next-generation mobile network 5G has already started hitting the market and will continue to expand worldwide, by providing everything (e.g., massive internet of things, camera drones, robotics, etc.) from superfast bandwidth speeds, to ultra-low latency, and to ultra-high reliability  \cite{lgy2}. 
5G will efficiently support three generic services, i.e., enhanced mobile broadband (eMBB), massive machine-type communications (mMTC), and ultra-reliable low-latency communications (uRLLC) \cite{haibo2}. 
The ever-increasing traffic demand, the rapid growth of the number of network applications and services, and the blossoming new service
requirements, are the driving forces for the complexity of mobile systems. This complexity resides not only in the sheer number of components and interactions, but also in the operations on multiple levels and time-scales \cite{shafin2020artificial}. The daunting complexity of mobile systems appears as a major hurdle for large-scale modeling efforts, which try to cognize user equipment, radio environment, and mobile network elements, and then apply it for network planning, service deployment, management functions, and resource scheduling. 
\par 
Currently, cognition is built on rules derived from system analysis and simulation with prior domain knowledge and experiences. For example, during LTE radio network planning, a general propagation model that doesn't incorporate the actual geographical information (terrain model) is applied to estimate coverage. As for capacity, link budget calculation, and theoretically assumed traffic are adopted to estimate cell size and capacity. After the cell has been deployed, it would operate and behave according to preprogrammed rules (protocols). In a nutshell, the mobile network is treated as a white-box, which requires an idealized abstraction and simplification of the underlying network. However, this method faces increasing challenges due to today's dynamic and diverse traffic, and the complexity of network architectures and resource structures. Besides, protocols cannot make intelligent decisions, and only react when a problem occurs. Consequently, conventional mobile networks are reactive rather than proactive \cite{lgy1}, e.g., transmission control protocol (TCP) just slows down the transfer speed when packets are lost and tries to slowly increase the speed until the next packet is lost. If it had knowledge of network status and condition before the transfer, it could optimize transfer speed and avoid packet loss.
\par 
Artificial intelligence (AI), with its emphasis on understanding how the human mind organizes and processes massive amounts of information, has made considerable achievements in the areas of computer vision \cite{zhang2020virtual, lgy3}, speech recognition, natural language processing, and audio recognition \cite{Chengnan}. Specifically, AI has produced results superior to human experts in strategic game systems (e.g., Deep Blue, AlphaGo, and AlphaStar, etc.), which demonstrates the potential ability in terms of cognition and decision.
Therefore, powered with mobile network big data (e.g., signaling data, user access data, connection data, etc.) and a plethora of computational power (mobile edge computing, and cloud computing), AI has the potential to cognize complex network environment, and arm it with the ability ``see through the future''. Then, proactive strategies and intelligent decisions can be taken to manage network resources, considering both user experiences and network performance. AI will revolutionize the current mobile network from the following aspects:
1) comprehensive cognition of user equipment, radio environment, and mobile network elements. With AI, it is possible to understand the user preferences and user traffic patterns, to construct a radio map \cite{9102392}, and to cognize the state of mobile network elements. 
2) intelligent decisions with orchestration between individual welfare and social welfare. AI can analyze and mine historical data to discover the reasons behind the past successes and failures, and then, exploit the extracted knowledge and experiences to properly allocate resources, guaranteeing end users with a high quality of experience (QoE).
3) self-adaption to system renovation and dynamics. By learning from past behaviors, and similar entities in the network, AI will equip a mobile network with the ability of self-learning and adaptive to the dynamics.
\par 
In this paper, we focus on the potential opportunity offered by harnessing AI into MN. Specifically, we first propose an AI-powered mobile network architecture, which has incorporated AI to the tasks of perception, cognition, and decision, under the emerging paradigm of mobile edge computing. Then, we investigate  challenges and discuss potential solutions, from the aspects of cognition complexity, decisions with high dimensional action space, and self-adaption to system dynamics. Finally, we exploit deep learning to  directly map the state and condition of mobile networks to the perceived QoS, thus helping network operators making efficient and intelligent admission control strategies for QoS provision. Our experiments on the real-world data  collected from a China telecom operator demonstrate the 
effectiveness and advantages of the proposed method, which involves $31261$ users over $77$ stations within $5$ days. 

\par

The remainder of this paper is organized as follows: Section II introduces the architecture of AI-powered mobile network. Section III focuses on the challenges and corresponding solutions are proposed in Section IV. Finally, Section V presents a case study that applies deep learning approach to map the state of MN to the perceived QoS.

\section{The Architecture of AI Powered Mobile Network}
\begin{figure*}
	\centering
	\includegraphics[width=0.94\textwidth,angle=0]{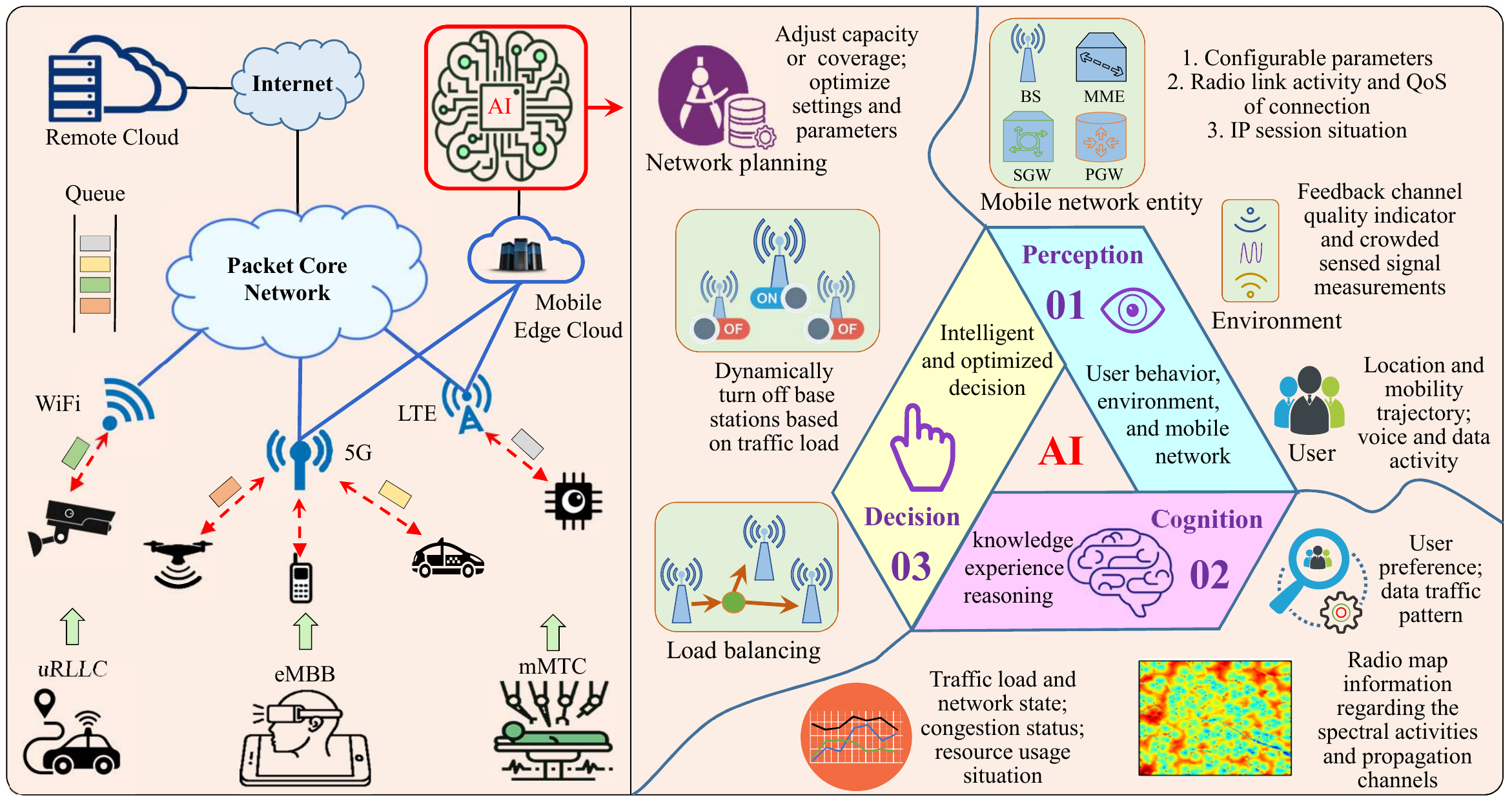}
	\caption{The proposed architecture for AI-powered mobile network, which consists of perception, cognition, and decision, forming a closed-form control
		system. These three parts are located in the mobile edge computing center, which can reduce latency, lessen the burden of the backbone network, and provide requisites (big data and computing resource) to releasing the power of AI. Consequently, the mobile network is equipped with the ability of real-time analytics, better customer experience, and lower operational expenditures. }
	\label{fig:architecture}
\end{figure*}
Computing power and big data are essential ingredients in releasing the power of AI. Mobile edge computing, which is a key technology toward 5G, distributes computing, and storage resources anywhere along the continuum from a remote cloud down to things \cite{9052677}. Meanwhile, a significant amount of data has been generated and accumulated every second at the mobile network system. With careful orchestration of these two ingredients, we propose the architecture of an AI-powered mobile network, 
which is shown in \reffig{fig:architecture}. Here, the Packet Core Network can either be the Evolved Packet System (EPS) in 4G and Next-Generation Core for 5G, which manages all operations needed to transfer voice and data to and from external networks. As for the base station (BS), it could be either the eNodeB in 4G and the next-generation NodeB (gNB) in 5G. The proposed architecture exploits the resources (computing, storage, and big data) available to mobile network operators (MNO), to ameliorate the planning, optimization, operation, and management of  MN. The proposed architecture consists of three parts, i.e., perception, cognition, and decision, which form a closed-loop system. 
\subsection{Perception}
Massive data that represent the condition of MN can be collected by mobile network operators, which we divide into three categories, i.e.,  mobile terminal data (MTD),  radio access network data (RAND), and packet core network data (PCND). 
MTD are collected by user equipment, consisting of service characteristics, device information, and wireless network parameters, cell ID, signal strength, download/upload rate, etc. The subscriber profiles and geographic information data are also included in MTD, which are beneficial for understanding and exploiting the features of mobile data from a social network perspective \cite{9141213}. These data depict the state and condition of the underlying user, and radio propagation environment.
RAND capture signaling events concerning any radio resource control (RRC) operation. This allows recording fine-grained state changes of each device, and thus detects device network attachment and detachment operations, start and conclusion of sessions, related to any call, texting, or data transfer activity. Moreover, they allow collecting performance indicators on data transmission, such as the uplink and downlink throughput experienced by the user equipment. 
The signaling data can indicate the aggregate-level status and condition of the mobile system, e.g., RRC connection success rate, EPS radio access bearer connection success rate. Besides, RAND also include the settings and configuration data, i.e., location, coverage and capacity, frequency band, maximum transmission power, antenna bearing, which are necessary parameters to initialize a mobile network so that it can operate as designed.
PCND collect information include the IP session start and end time, device and user identifiers, traffic volume, type of service (i.e., web, email, streaming audio/video). PCND can also be used to infer user preference and user attributes, by monitoring the session length (number of pages a user clicks through) and abandonment rate  (if a user leaves the website after visiting the landing page) \cite{keller2020quality}. Besides, the feedback channel quality is a good perception of the wireless environment.

\subsection{Cognition}
The massive data offer tremendous opportunities in learning and understanding user equipment, radio propagation environment, and mobile network elements (i.e., EPS and E-UTRAN in LTE). User preferences, user mobility, and user traffic behaviors are among the most important features that affect the performance of the cellular network. User preferences are associated with the QoE perceived by the user. The QoE is of significant importance, since the ultimate goal of the cellular network is to improve the overall perceived QoE. User mobility, and user traffic behavior influence the traffic load of each base station, thus affecting the overall performance. This user-related knowledge can be learned and mined from RAND and MTD \cite{keller2020quality}.
Radio propagation environment account for the propagation mechanisms of electromagnetic waves on a wireless link, which is affected by environment dynamics, obstacles distribution, noise, weather, etc. Such environment information can be leveraged to not only improve the performance of existing wireless services (e.g., via proactive resource provisioning) but also enable new applications, such as wireless spectrum surveillance. 
Almost every radio link activity is associated with a signaling exchange inside the mobile network system. The signaling data equip us with real-time awareness of the traffic load, resource utilization, and network state. By associating these data to mobile users, the relationship between user traffic and mobile network state can be properly learned. 

\subsection{Decision}
With the extracted knowledge of the user equipment, radio environment, and mobile network elements, more optimized network planning and intelligent scheduling strategies can be achieved. However, the strategies involve actions within different layers, granularity, and even mixed strategies. For example, based on the analysis of historical traffic load, we could plan the network by adjusting the capacity and coverage of a cell site (i.e., increase coverage, or deploy more small base stations). Besides, we could also optimize the antenna direction and down tilts, mobility (handover and cell re-selection) parameters for each cell, based on learned traffic patterns. 
Except for long-term network planning, there exist many scheduling strategies that deal with real-time situations. When there is a slight traffic load, we could choose to reduce energy consumption by dynamically switching
off base stations. With the traffic load increased, various load balancing scheme can be applied, e.g., offloading a part of traffic to a neighboring cell, or other networks (WiFi, 3G, etc.). Continuously increased traffic load would lead to traffic congestion, and this can be detected via learning the relationship between traffic load and the state of the LTE system. To handle this situation, we could reject a part of low-priority data requests in advance to prevent the congestion from affecting all the users within its coverage.

\section{Challenges}

\subsection{Complexity and correlation in cognition}
The mobile network system is affected by the performance of individual nodes and protocols through
which information travels, due to the use of packet switching. Consequently, joint cognition of user equipment, radio environment, and mobile network elements are essential to make intelligent and optimized decisions. However, the cognition is not only complicated but also significantly correlated. For example, cognition of user includes both per-user level (mobile user viewpoint) and aggregate level traffic (mobile operator viewpoint) \cite{wey20205g}. Per-user level traffic is related to user preference, mobility, and varying traffic demands. The aggregation of these data requests exhibits traffic patterns, e.g., spatio-temporal dynamics, anomalous behaviors induced by particular social events. 
The radio environment serves as a bridge for exchanging data between the user equipment and mobile network elements, which is characterized by a radio map (a continuous function of location, frequency, and time) \cite{RadioMaps}. It contains rich and useful information regarding the spectral activities and propagation channels in wireless networks. Constructing the map is challenging since it requires real-time measurements and constantly updated. Mobile network elements manage and allocate the resources through sophisticated software and protocols, considering the traffic demands of all users and the radio environment. Therefore, the user equipment, radio environment, and mobile network elements are heavily correlated. 
\par 
The complexity and correlations pose tremendous challenges in smoothing and optimizing network performance. Take the perceived QoE of a certain user as an example. It is related to the user, radio environment, and mobile network elements. Besides, it is also related to both user characteristics (different perceived QoE even with the same QoS) and user traffic (different services require different QoS). The radio environment affects the transmission rate of the connection. The mobile network elements, which forwards the packets to the Internet, would also influence the QoS of the connections. Furthermore, the situation beyond the mobile network can also deteriorate the QoS. Therefore, comprehensive cognition of the mobile network system is challenging, since it not only complex but also heavily correlated.
\subsection{High dimensional state/action spaces in decision}
The future mobile network must empower itself with more advanced technologies to support a tremendous number of applications, meet the significantly contrasting QoS requirements (bandwidth, latency, and reliability), and improve the spectrum efficiency. Therefore, operators necessarily consider decisions over multiple dimensions and granularities. At first,
networks are becoming increasingly heterogeneous, and user devices can be served by more than one technology, such as 2G, 3G, 4G, Wi-Fi, etc. Within each technical domain, operators have options to combine multiple layers of cells (i.e., base station, small cell, femtocell), and various radio beams can be organized to better serve expected data requests. Furthermore, in each connection, there exist multiple technologies (carrier aggregation, beamforming, device to device communication, non-orthogonal multiple access, etc.), offering tremendous opportunities to exploit resources in time, spectrum, space, and power domain. To make full use of these resources and better improve the perceived QoE, the user characteristics and radio environment have to be taken into account. For example, millimeter wave, which is a key technology enabler for 5G and can be used for high-speed wireless broadband communications, travels by line of sight, since its short wavelengths (wavelengths from 10 millimeters to 1 millimeter) can be blocked by physical objects like buildings and trees. Additionally, base stations can be turned on or off dynamically to manage load or interference, thus reducing capital expenditures and operational expenditures. Furthermore, the decisions among neighboring cells are dependent, e.g., interference to users in neighboring cells, offloading from overload cell to underload cell. Therefore, it is significantly challenging to make intelligent and optimized decisions due to the high dimensional state/action space.

\subsection{Self-adaption to the evolving network dynamics}
Driven by the prosperity and flourishing of information technology, and by the proliferation of connected devices, the network is always evolving. The evolving dynamics can be induced by users, the radio environment, mobile network elements, events, etc. Born in the early 1980s, the first generation of wireless cellular technology 1G supports voice calls only, applying analog technology (poor battery life and voice quality, little security, and prone to dropped calls). The introduction of 3G networks in 1998 ushers in more data-demanding ways such as for video calling and mobile internet access. The imminent arrival of 5G promises significantly faster data rates, higher connection density, much lower latency, and energy savings, among other improvements. The evolution of the mobile network has significantly enriched the scope and horizon for mobile users, giving birth to a plethora of applications and services with distinct QoS requirements. Consequently, user traffic and content requirements variation is constantly evolving \cite{fenglyu}, driven by both the evolution of technology and user characteristics. The radio environment can be influenced by the construction, the weather, and moving objects. Besides, national festivities (e.g., Thanksgiving, Christmas, New Year's Eve, Easter, etc.), events (e.g., the Olympic games, sports, solo vocal concert) would significantly affect the traffic content and user patterns, thus imposing influence on network performance. These dynamics could fail the deployed strategies or rules. To handle these dynamics, self-adoptive and agile strategies and decisions shall be taken to better meet communication requirements.

\section{Solutions}
AI makes it possible for machines to accomplish specific tasks by processing large amounts of data and recognizing patterns in the data. Thanks to increased data volumes, advanced algorithms, and improvements in computing power and storage, AI has achieved considerable improvements in the area of computer vision, natural language processing, bioinformatics, etc. AI can be loosely interpreted to mean incorporating human intelligence to machines, which consists of everything from Good Old-Fashioned AI (GOFAI) all the way to futuristic technologies such as deep learning. GOFAI, which is also known as symbolic AI, can identify the influences and reasons of a certain event in the mobile network, e.g., determine the influence when a base station is broken down. Machine learning, as a subset of AI techniques that do not require human intervention to make certain changes, was coined by Arthur Samuel in 1959.
Deep learning (also known as deep structured learning or hierarchical learning),
being a subset of machine learning, is inspired by the information processing patterns found in the human brain and consists of thousands or even millions of simple processing nodes that are densely interconnected. 
Deep learning, usually with more (millions of) parameters, can represent much more complicated functions, thus standing out as a powerful approach for image classification, natural language processing, etc. Besides, deep learning methods obviate feature engineering by translating the data into compact intermediate representations, and derive layered structures that remove redundancy in representation. 
\par

\begin{figure}
	\centering
	\includegraphics[width=0.48\textwidth,angle=0]{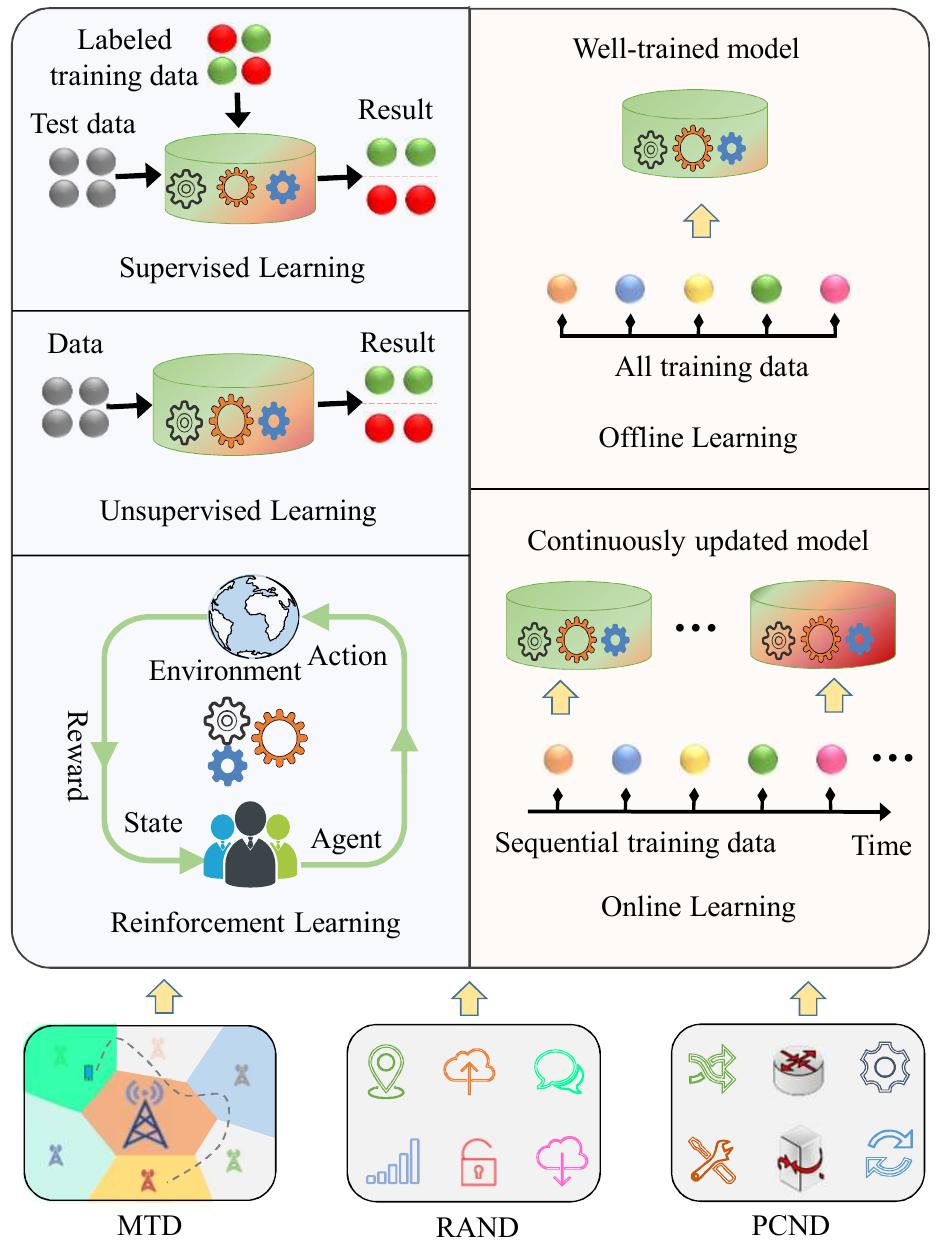}
	\caption{ Machine learning algorithms are divided into three broad categories, i.e., SL, UL, and RL, respectively, depending on the nature of the signal or feedback available to the learning system. They can also be categorized into ONL and OFL based on the static dataset or sequentially generated dataset. Each model has different requirements on the type of data, and can be applied to different tasks. }
	\label{fig:machinealgs}
\end{figure}

Machine learning empowers machines with the ability to learn without being explicitly programmed to perform specific tasks. With the availability of 
massive volume data, machine learning is widely hugged, since it can learn from data, identify patterns, and make decisions with minimal human intervention. Machine learning can be divided into three classes, i.e., supervised learning (SL), unsupervised learning (UL), and reinforcement learning (RL), respectively. In a SL model, the algorithm learns on a labeled dataset, providing an answer key that the algorithm can use to evaluate its accuracy on the training data. An unsupervised model, in contrast, provides unlabeled data that the algorithm tries to make sense of by extracting features and patterns on its own. As for the RL model, the agent relies both on learning from past feedback and exploration of new tactics that may present a larger payoff, to make its decisions \cite{lgy2}. Machine learning can also be divided into offline learning (OFL) and online learning (ONL), based on whether to change their approximation of the target function during the training. The whole training data of an OFL model must be available at the time of model training. Only when training is completed can the model be used for predicting. ONL is a very special kind of learning approach, which processes data sequentially. They produce a model and put it in operation without having the complete training dataset available at the beginning. The model is continuously updated during operation as more training data arrives, which is shown in \reffig{fig:machinealgs}. 

\subsection{Supervised learning and unsupervised learning contribute to better cognition}
\par 
To perform a task, a SL model requires a set of training examples, and each sample in the dataset consists of an input object (typically a vector) and the desired output value (ground truth). A SL algorithm analyzes the training data and produces an inferred function, which can be used for mapping new examples. SL problems can be further grouped into regression (e.g., linear regression, logistic regression) and classification (e.g., support vector machines, neural networks) problems. It can be adopted to cognize the traffic load of a mobile network system, by ingeniously applying QoS changes as labels. The traffic load of a mobile network significantly affects the QoS of connections, since users have to share the network resources. However, the traffic load is correlated to many values in key performance indicators (KPI) of mobile systems, e.g., physical resource blocks utility rate, physical downlink control channel utility rate, physical random access channel utility rate, radio resource control success rate, etc. As for the label, we could treat the average perceived QoS as the indicator for the condition of traffic load, since more traffic load might lead to traffic congestion, which significantly degrades QoS. Then, we could apply a classification algorithm over KPI to identify the traffic load condition. Based on this traffic load, strategies that prevent traffic congestion can be automatically adapted, which will be elaborated in \ref{usercase}.
\par
The training data for UL does not possess prior labels. Consequently, the UL agent has to depend on its capability to learn the embedded structure or pattern in data. UL is very popular, since it does not require a labeled or classified dataset, which is very time-consuming and labor-intensive. Clustering (e.g., K-means, hierarchical clustering, OPTICS algorithm, etc.) and dimensionality reduction (e.g., principal component analysis, singular value decomposition, etc.) are two types of UL approaches. Clustering can be applied to understand the aggregate-level traffic activity. For example, base stations that are associated with different types of regions (e.g, work, residential, hybrid, nightlife, and leisure) exhibit significantly different traffic profiles. For the work areas, the activity takes place mainly during weekdays, especially
during working hours, while weekend activity is almost nonexistent. However, for leisure areas, activity on weekends is more than twice that of weekdays, and this activity is focused during daylight hours. Exploiting clustering algorithms, we could divide base stations into different clusters, which are characterized by aggregating traffic activities. We could optimize the strategies for each cluster according to the traffic characteristics.

\begin{figure}
	\centering
	\includegraphics[width=0.49\textwidth,angle=0]{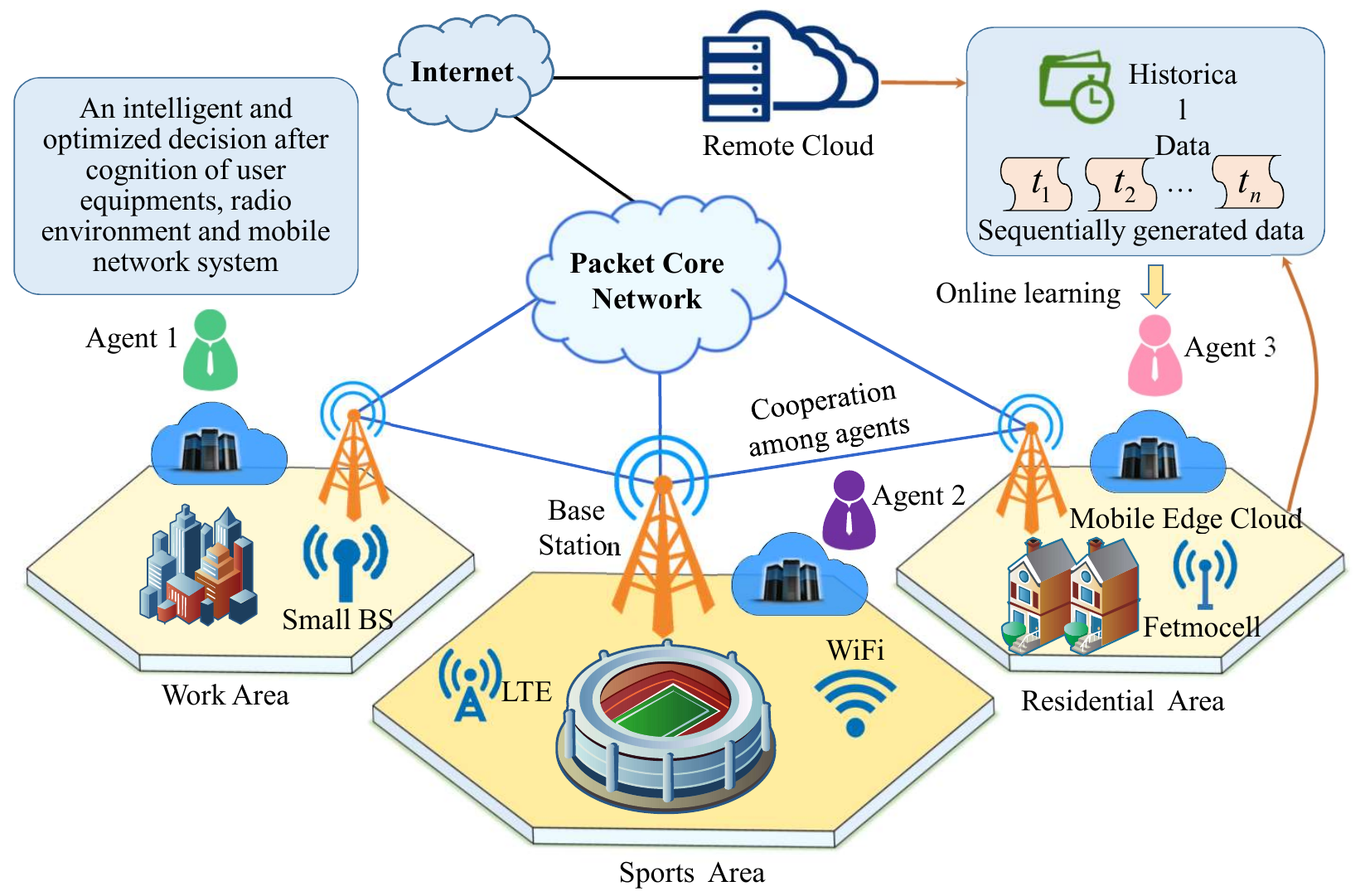}
	\caption{Online learning equips each base station with self-adaption ability. }
	\label{fig:edgeCloud}
\end{figure}

\subsection{Deep reinforcement learning for intelligent and optimized decisions}
RL is the area of machine learning that deals with sequential decision-making \cite{haibo1}, which can be formulated as a Markov decision process. At first, an agent, with a given environment state $s_t$ from the state space $\mathcal{S}$, attempts to take an action $a_t$ in the state space $\mathcal{A}$. Then, the agent receives a reward and transits to the next state $s_{t+1} \in \mathcal{S}$. In an episodic problem, this process continues until the agent reaches a terminal state, and then it restarts. The agent aims to maximize the accumulated reward of the long term return from each state. An RL agent needs a trade-off between the exploration of uncertain policies and exploitation of the current best policy, which is a fundamental dilemma in RL. Deep reinforcement learning, which is a combination of RL and deep learning, equips RL with higher generalization ability. Consequently, deep RL can model high dimensional state/action spaces in decision. For example, a deep RL based AlphaGo, which has defeated a Go world champion, has shown its potentials in modeling high dimensional state and action space, since Go is a game of profound complexity and there is an astonishing $10^{170}$ possible board configurations. Furthermore, games like StarCraft II that are harder for computers to play than board games like chess or Go, have also been conquered by deep RL based models.
\par 
Deep RL can be applied to learn intelligent resource allocation and scheduling decisions. We model the state of the users (application type, QoS requirements), radio map (channel state information), and KPI in mobile network as state space $\mathcal{S}$. The action space $\mathcal{A}$ consists of resource allocation (allocated resource blocks, power) and scheduling (access to which cells, D2D or beamforming techniques). The goal of deep RL is to learn a policy $\pi:\mathcal{S}\rightarrow \mathcal{A} $, maximizing long term rewards, which can be defined as the average throughput, perceived QoS, or users' satisfaction (QoE). 

\subsection{Online learning to build a self-adaptive system}
An ONL agent makes online, real-time decisions and continuously improve performance with the sequential arrival of data. It has a significant difference over OFL (also known as batch learning), which is trained over the entire dataset. ONL algorithms are widely applied for advertisement placement, movie recommendation, and node or link prediction in evolving networks. It can help the model to adapt to dynamics in the users, radio environment, and mobile network system, aiming at dynamically optimizing the quality of experience. The dynamics can be divided into slow variation (e.g., user behavior and aggregate-level traffic, radio environment) and drastic variation (e.g., mobile network system, events). The former one alters very slowly, such as  traffic patterns, which reveals differences during daytime and night, or weekdays and weekends. Sequentially generated data can be applied to update the model, and each cell is associated with a model, as shown in \reffig{fig:edgeCloud}. However, drastic variation changes sharply due to newly installed cells or national festivities or events. For these variations, we require the experience and knowledge from historical data or a similar area. For example, a solo vocal concert would gather a mountain of people in a certain area within a short duration. The traffic pattern within this area has changed significantly, and the mobile network system has to be carefully configured. To achieve this, we should first search historical events or data in the remote cloud, and extract the knowledge for such events. Then, based on the radio environment, and estimated traffic, we can reconfigure the mobile network in this area or deploy a temporary access point to offload part of traffic.

\section{A User Case: QoS provision from network operator point view}
\label{usercase}
Network operators aim at providing better QoS connections to users while admitting as many users as possible. However, users within a cell share the available resources, and more users could potentially lead to traffic congestion, which significantly influences the perceived QoS of all users. Therefore, the network operator should first cognize the condition and state of the cell, and then deal with (admission control) the connection requests based on network condition, thus preventing network congestion from happening. To achieve this goal, based on real-world data collected from a telecommunication operator in China, we propose a deep learning based model that integrates cognition with decision, to provision QoS. Our proposed model directly maps the condition and state of the mobile network systems to admission control strategies. The well-trained of such model can be applied to help the cell in making more intelligent and optimized decisions, thus preventing it from traffic congestion.
\subsection{Mobile Network Dataset}
\par
\begin{figure}
	\centering
	\includegraphics[width=0.48\textwidth,angle=0]{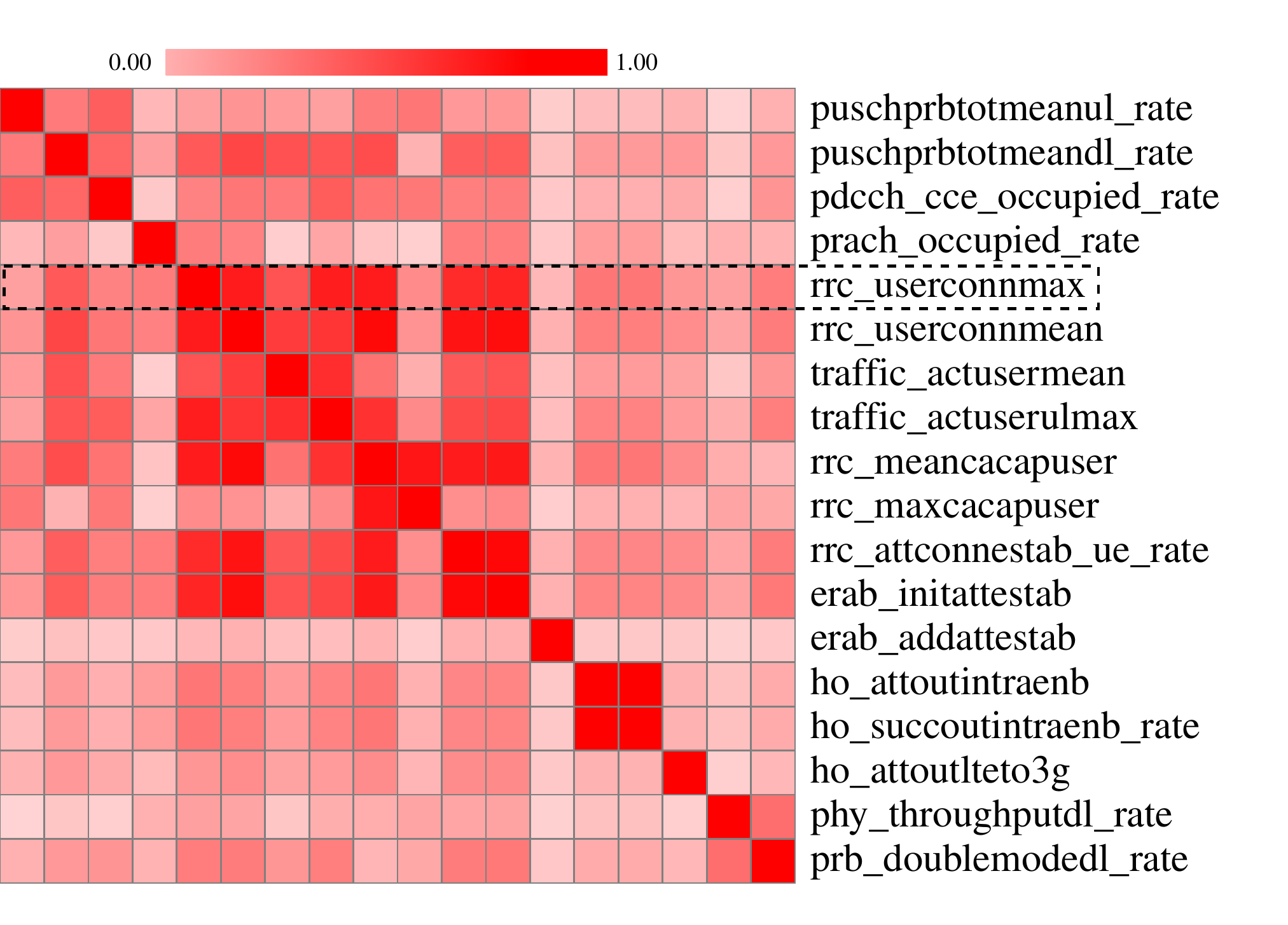}
	\caption{Heat map of the KPI Pearson correlation coefficients. }
	\label{fig:heatmap}
\end{figure}

The real-world data are collected from a China telecom operator, containing data from $31261$ users, over $77$ stations within $5$ days. It consists of $4$ parts, i.e., KPI, measure reports (MR), traffic detail records (TDR), and data monitored by user equipment (DMU).
KPI records the performance values of components in the mobile network, and is already defined in 3GPP TR 32.814.
KPI helps network operators to monitor and optimize the radio network performance, providing better subscriber quality, and achieving a higher utilization ratio of the network resources.
KPI includes radio KPI and transport KPI, e.g.,
RRC connection establishment success rate (obtained by the number of all successful RRC establishments divided by the total number of attempted RRC establishments and abbreviated as $rrc\_attconnestab\_ue\_rate$ in \reffig{fig:heatmap}.),
soft handover success rate (obtained by the number of successful radio link additions divided by the total number of radio link
addition attempts and abbreviated as $ho\_succoutintraenb\_rate$ in \reffig{fig:heatmap}.).
MR is a statistical data report that contains information about channel quality, which is generated by measuring the signal quality, e.g., average signal to interference plus noise ratio, weak coverage range, average reference signal receiving quality, etc. 
TDR records the event and the type for all data traffic, e.g., the traffic is served by which cell, and at which time, and the traffic characteristics.
DMU records the real-time QoS (delay, jitter, packet loss, etc.) of each connection, which is reported by user equipments.
\par 
\subsection{Correlation and Hidden Patterns}
KPI parameters  are dependent. For a KPI parameter $u$, the value at $t$-th timestamp is denoted by $x^u_t$. Therefore, the values of $5$ days forms a vector $X^u=[x^u_0,x^u_1,\cdots, x^u_T]$. For two KPI parameters $u$ and $v$, Pearson correlation coefficient (PCC) of $X_u$ and $X_v$ is denoted as $\rho_{u,v} = \frac{cov(X_u,X_v)}{\sigma_{X_u}\sigma_{X_v}}$, where $cov(X_u,X_v)$ is the covariance and $\sigma_{X_u}$ is the standard deviation of $X_u$. We have a set of KPI parameters and adopt absolute PCC as indicator for correlations, i.e.,$|\rho_{u,v}|$. A matrix is applied to represent the correlations of these parameters, as shown in \reffig{fig:heatmap}. As can be seen from this figure, the KPI parameters are heavily correlated, e.g., in the fifth row of \reffig{fig:heatmap}, $rrc\_userconnmax$ is correlated with several other parameters. Due to this correlations, we could not depend on a single parameter to identify the state and condition of a cell, e.g, applying uplink physical resource block average usage rate (abbreviated as $puschprbtotmeanul\_rate$ in \reffig{fig:heatmap}) as the indicator for congestion evaluation, since it is related to other KPIs. If all KPI parameters are considered, a more accurate state and condition can be obtained, and more accurate performance (e.g., congestion evaluation) can be achieved. 
\par 
Besides, there exist some hidden features and patterns within these data. For example, a decreasing RRC connection establishment success rate could potentially degrade perceived QoS. We propose to discover these patterns through deep learning technologies. 
\subsection{QoS Provision Based on Deep Learning}
\par 
Our objective is to design admission control strategies for QoS provision. We achieve it by identifying the state and condition of the mobile network system that will lead to poor performance in QoS. Then, corresponding strategies could be applied to provision QoS, e.g., early reject low-priority data requests. The KPI and MR data can well represent the state and condition of the mobile network system, while the DMU and TDR data contain the information regarding traffic connection, including the perceived QoS. The change in average perceived QoS is treated as the learning target. In the implementation, we treat such problem as a binary classification problem, where  KPI and MR of a cell, and characteristics of the data connections are treated as input and the change of average perceived QoS as the output (QoS improvement or deterioration). These input-output pairs consist of labeled data. Among these data, $70\%$ is treated as the training dataset, $20\%$ as the test dataset, and the rest as the validation dataset. 

\par
\begin{figure}
	\centering
	\includegraphics[width=0.459\textwidth,angle=0]{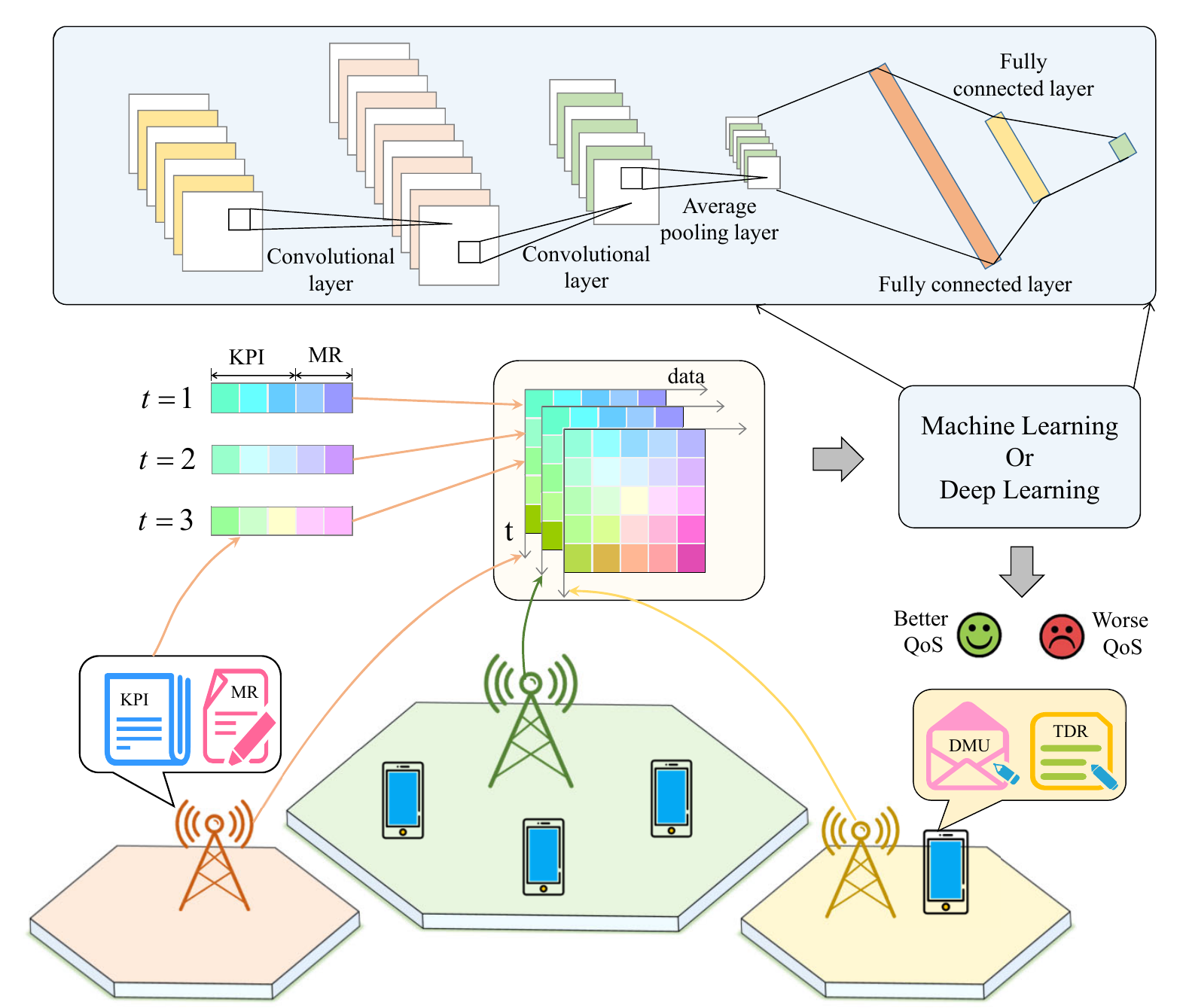}
	\caption{We propose a deep learning algorithm that directly map the state and condition of mobile network system to the  perceived QoS. The proposed algorithm can identify hidden features and patterns within MN, i.e., the state of mobile network system that could lead to traffic congestion. }
	\label{fig:approach}
\end{figure}

 
To capture the spatio-temporal dependences, the KPI parameters across different timestamps and adjacent cells are treated as a three-dimensional tensor, as shown in \reffig{fig:approach}. Convolution neural networks (CNN), which take advantage of the hierarchical patterns in data and assemble more complex patterns using smaller and simpler patterns, are applied to extract features and then map them to the ground truth labels. The adopted CNN model consists of two convolutional layers, an average pooling layer, and two fully connected layers, and is shown in \reffig{fig:approach}. Furthermore, we have adopted several machine learning algorithms, which are k-nearest neighbors (KNN), support vector machine (SVM), decision tree (DT), and light gradient boosting machine (LGB).

\par
\begin{figure} 
	\centering 
	\subfigure[] { \label{fig:AlgsComparison} 
		\includegraphics[width=0.4\textwidth,angle=0]{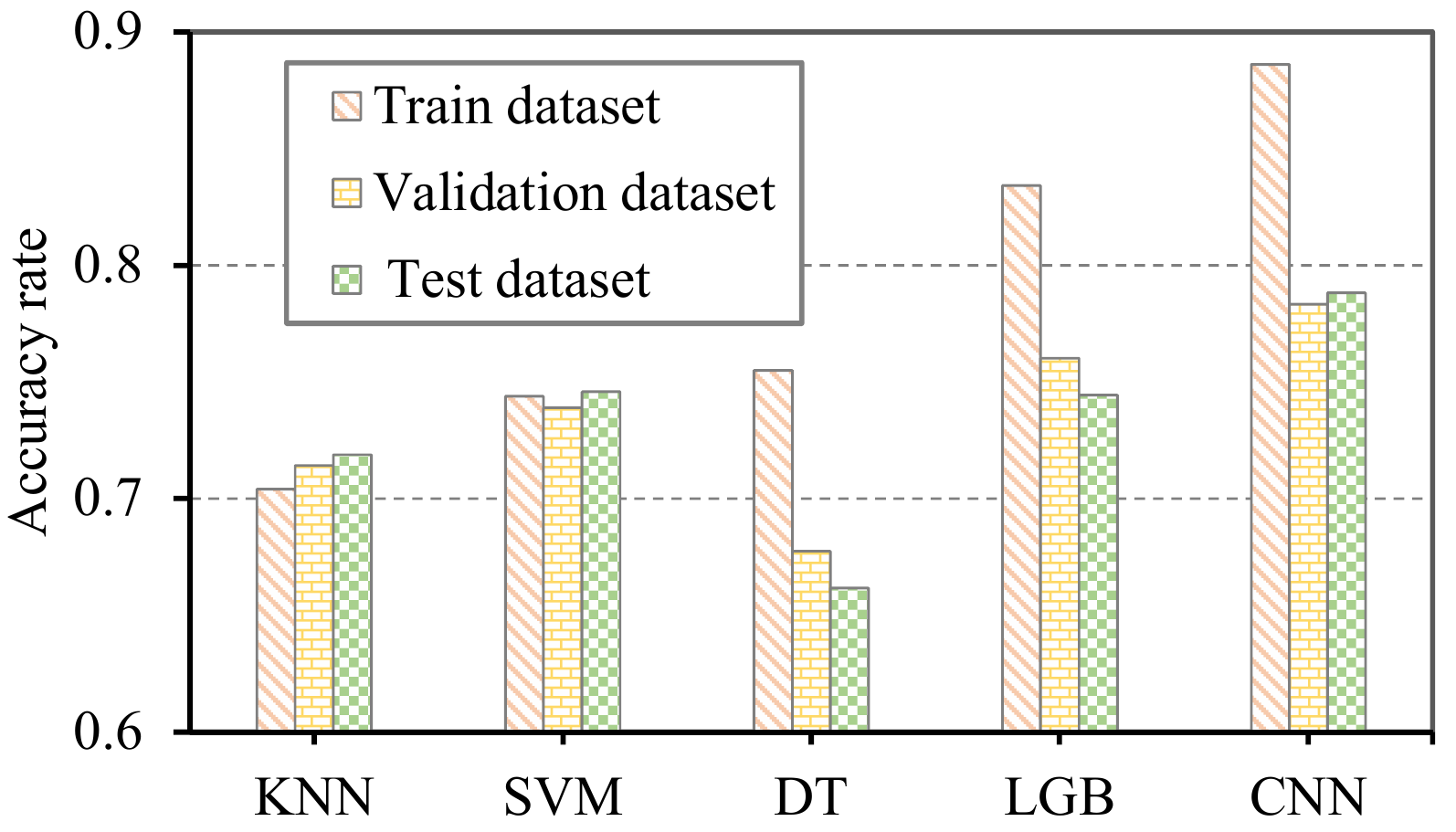}
	} 
	\subfigure[ ] { \label{fig:QoScomparison} 
		\includegraphics[width=0.45\textwidth,angle=0]{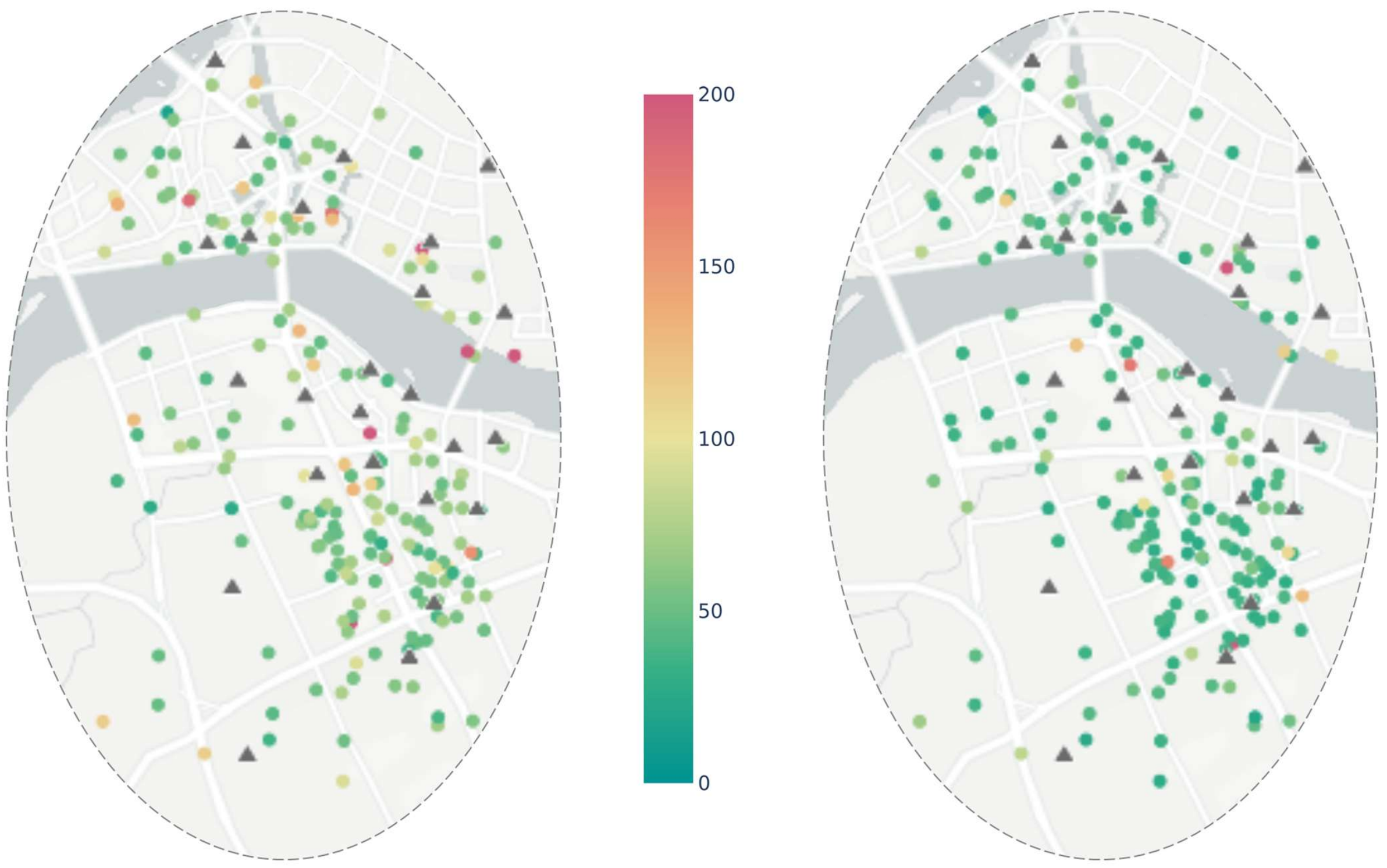}
	} 
	
	\caption{ \hl{We propose a deep learning algorithm CNN that directly map the state and condition of the mobile network system to perceived QoS. The proposed algorithm can identify hidden features and patterns within MN, i.e., the state of the mobile network system that could lead to traffic congestion. (a) shows the performance of CNN as compared with several machine learning baselines, i.e., KNN, SVM, DT, and LGB. Our proposed  CNN achieves the best performance. (b) illustrates the perceived delay without (left) and with (right) CNN  to provision QoS. The color bar maps the delay (millisecond) to different colors, indicating our proposed CNN  could potentially decrease the perceived delay.
	} } 
	
\end{figure}

 
The experiments results are shown in \reffig{fig:AlgsComparison}. For each traditional machine learning method, we have deliberately chosen the features that could lead to better performance, i.e.,  with careful feature engineering. A different number of input features is evaluated and the performance of the best is shown in \reffig{fig:AlgsComparison}.  Traditional machine learning methods achieve a poor performance, mainly due to: 1) the model is shallow, which can't handle very complex function approximation and learn the non-linear and complex relationship between the state of MN and admission control decision; 2) the model is with little trainable variables, which is prone to be over-fitting. CNN achieves the best performance, with $78.84\%$ accuracy on the test dataset. CNN is preferred in this case, not only for its better performance, but also for its simplicity, i.e., without any feature engineering. This well-trained CNN  model associates the state of the mobile network system with the perceived QoS of users, and can be applied to help the operator in deciding on how to better allocate resources considering the state of the mobile network system. 
\par 
\reffig{fig:QoScomparison} visualize the performance enhancements by applying the CNN model for QoS provision, where the triangle means base stations, and the circle stands for a user equipment.  The left figure of \reffig{fig:QoScomparison} expresses the originally perceived delay, while the right shows the perceived delay applying our proposed CNN model for QoS provision. The results verify that the proposed CNN model could decrease the perceived delay and provide better QoS to users.

\section{Conclusion and Future Work}
AI has offered tremendous opportunities for network operators to plan, deploy, manage, and maintain the network, which can not only better satisfy user requirements but also significantly reduce capital expenditures and operational expenditures. This paper first proposes an AI-powered architecture. Then, challenges and potential solutions regarding the application of AI in the mobile network are investigated. Finally, we have proposed a deep learning approach that directly maps the state of the mobile network to perceived QoS and demonstrated its effectiveness on real-world data collected from a China telecom operator. To fully leverage the power of AI in the mobile network, further research directions are discussed as follows:
\par 
\textbf{Integration of terminal intelligence, edge intelligence, and cloud intelligence ---} With the development of information and communications technology, the terminal (i.e., mobile phone, wearable devices, automated vehicles, etc.) are evolved with higher intelligence. These terminals are aware of the usage patterns and preferences of the user, thus can make more intelligent decisions in advance. For example, they could automatically cache videos or news in advance when the network condition is good, taking the user preference and network condition into account. Integration of the intelligence from terminal, edge, and cloud could significantly improve the efficiency of the mobile network system.
\par
\textbf{Privacy and security in utilizing the massive data ---} Although machine learning, especially deep learning, has shown enriched performance results in a variety of fields, the model can easily lead to mispredictions or misclassifications due to malicious external influences. Besides, the massive data in the mobile network show promising potential in the areas of social networks, geography, and urban planning. Therefore, it is challenging to protect privacy in traffic data, since it contains sensible information on individual subscribers. 
\par
\textbf{Advanced technologies and platforms to exploit and manage the massive data ---} 
Users and mobile network entities are always generating data, which distributes from the things, to the edge and the remote cloud. Due to its enormous quantity, research challenges range from data collection, data storage and transmission, data transformation and processing, and finally data usage. Advanced technologies and platforms shall be carefully orchestrated to efficiently store, harness, and disseminate these massive data.

%
%

%
%

%

\bibliographystyle{IEEEtran}
\bibliography{4G_QoS}

\end{document}